\documentclass[journal,review,submit,moreauthors,pdftex,applsci]{mdpi}
\usepackage{color,soul}
\firstpage{1} 
\pubvolume{xx}
\issuenum{1}
\articlenumber{5}
\pubyear{2020}
\copyrightyear{2020}
\history{Received: date; Accepted: date; Published: date}

\Title{Fully superconducting Josephson bolometers for gigahertz astronomy}

\Author{Federico Paolucci $^{1}$, Nadia Ligato$^{1}$, Gaia Germanese$^{1,2}$, Vittorio Buccheri$^{1,3}$, and Francesco Giazotto $^{1}$}

\AuthorNames{Federico Paolucci, Nadia Ligato, Gaia Germanese, Vittorio Buccheri, and Francesco Giazotto}

\address{$^{1}$ \quad NEST, Istituto Nanoscienze-CNR and Scuola Normale Superiore, I-56127 Pisa, Italy\\
$^{2}$ \quad Dipartimento di Fisica dell'Universit\`a di Pisa, Largo Pontecorvo 3, I-56127 Pisa, Italy\\
$^{3}$ \quad INFN Sezione di Pisa, Largo Bruno Pontecorvo, 3, I-56127 Pisa, Italy\\}

\corres{Correspondence: F.P. federico.paolucci@nano.cnr.it; F.G. francesco.giazotto@sns.it}

\abstract{The origin and the evolution of the universe are concealed in the evanescent diffuse extragalactic background radiation (DEBRA). To reveal these signals, the development of innovative ultra-sensitive bolometers operating in the gigahertz band is required. Here, we review the design and experimental realization of two bias-current-tunable sensors based on one dimensional fully superconducting Josephson junctions: the nanoscale transition edge sensor (nano-TES) and the Josephson escape sensor (JES). In particular, we cover the theoretical basis of the sensors operation, the device fabrication, their experimental electronic and thermal characterization, and the deduced detection performance. Indeed, the nano-TES promises a state-of-the-art noise equivalent power (NEP) of about $5 \times 10^{-20}$ W$/\sqrt{\text{Hz}}$, while the JES active region is expected to show an unprecedented NEP of the order of $10^{-25}$ W$/\sqrt{\text{Hz}}$. Therefore, the nano-TES and JES are strong candidates to push radio astronomy to the next level.}


\begin{document}


\section{Introduction}
\begin{figure}[t!]
\centering
\includegraphics[width=\columnwidth]{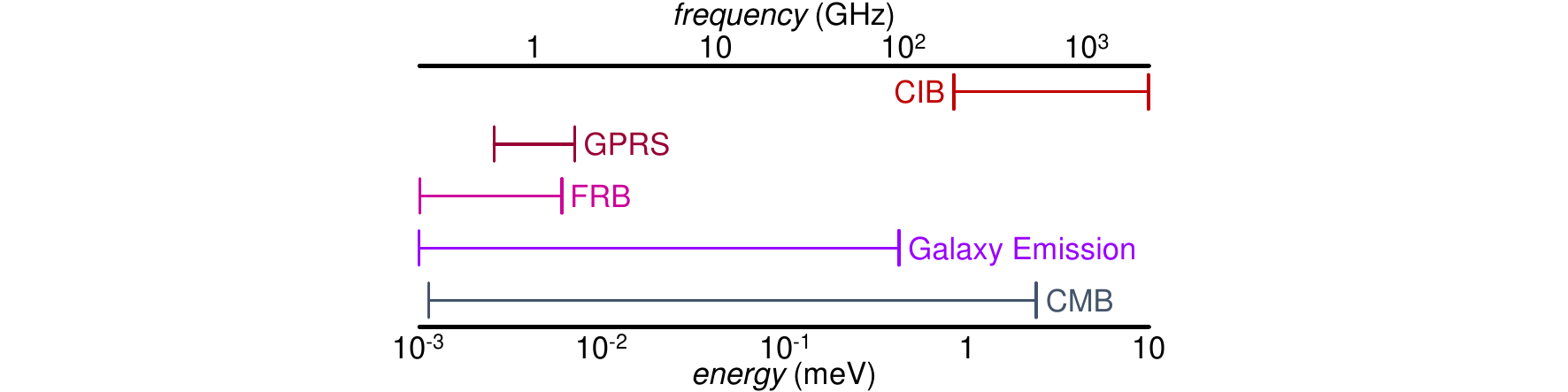}
\caption{Frequency and energy ranges relevant for different GHz astronomical sources, such as cosmic infrared background (CIB), gigahertz-peaked radio sources (GPRS), fast radio bursts (FRB), emission from galaxies, and cosmic microwave background (CMB).}
\label{FigAppl}
\end{figure}

The universe is permeated by radiation extending over almost all the electromagnetic spectrum, the so-called diffuse extragalactic background radiation (DEBRA) \cite{Primack}. Particular interest is posed on the portion of the DEBRA covering energies lower than 1 eV, since this range contains information regarding the formation and the evolution of the universe. Indeed, the temperature and the polarization of the cosmic microwave background (CMB) \cite{Seljak,Kamionkowski,Sironi}, ranging from 300 MHz to 630 GHz, reveal the early stages of the universe and represent the smoking gun proof of the Big Bang. Furthermore, the emission from stars/galaxies formation \cite{Tabatabaei2017} and the active galactic nuclei (AGNs) \cite{Fabian} gives rise to the cosmic infrared background (CIB) \cite{Cooray,Zemcov}, that shows $10\%$ of the power of the CMB. As a consequence, sensitive gigahertz photon detection is fundamental to shed light on different radioastronomy phenomena \cite{Rowan2009}, such as atomic vibrations in galaxy clusters \cite{Villaescusa-Navarro}, hydrogen atom emission in the galaxy clusters \cite{Armus}, radio burst sources \cite{Marcote2020}, comets \cite{Falchi1988}, gigahertz-peaked spectrum radio sources \cite{Odea1998}, and supermassive black holes \cite{Issaoun2019}. Figure \ref{FigAppl} resumes the frequency and energy range for selected radio-astronomy source in the GHz band.

State-of-the-art GHz/THz detectors for astronomy are the transition edge sensors (TESs) \citep{Irwin1995, Irwin2006, Karasik} and the kinetic inductance detectors (KIDs) \cite{Monfardini}, since they are extremely sensitive and robust. Indeed, TES-based bolometers reached a noise equivalent power (NEP) of the order of $10^{-19}$ $\mathrm{W/\sqrt{Hz}}$ \cite{Khosropanah}, while KIDs showed a NEP of about $10^{-18}$ $\mathrm{W/\sqrt{Hz}}$ \cite{Visser}. To push detection technology towards lower values of NEP, it is necessary a strong reduction of the thermal exchange of the sensor active region, i.e., the portion of the device absorbing the incident radiation, with all the other thermal sinks. To this end, miniaturized superconducting sensors have been realized \cite{Wei}. Furthermore, several hybrid nano-structures exploiting the Josephson effect \cite{Josephson} have been proposed and realized \cite{Solinas,Guarcello}. In particular, cold electron bolometers (CEBs) in normal metal/superconductor tunnel systems were experimentally demonstrated to show a $NEP\sim 3\times 10^{-18}$ $\mathrm{W/\sqrt{Hz}}$ \cite{kuzmin2019}, while gate-tunable CEBs can be realized by substituting the metal with silicon \cite{Brien} or graphene \cite{Vischi}. Furthermore, detectors based on proximity effect demonstrated a NEP of the order of $10^{-20}$ $\mathrm{W/\sqrt{Hz}}$ in superconductor/normal metal/superconductor (SNS) junctions \cite{Giazotto2008,Kokkoniemi} and $7\times10^{-19}$ $\mathrm{W/\sqrt{Hz}}$ in superconductor/graphene/superconductor (SGS) junctions \cite{Lee1}. Finally, an innovative sensor based on the exotic temperature-to-phase conversion (TPC) in a superconducting ring is expected to provide a $NEP\sim 10^{-23}$ $\mathrm{W/\sqrt{Hz}}$ \cite{Virtanen}.

Despite the high sensitivity, all these innovative bolometers do not fulfill the strict requirements for being employed in ultra-sensitive space telescopes. In fact, materials, fabrication procedures and structures need to be extensively tested to survive the extreme stress (vibrations, acceleration, radiation, etc.) typical of space missions. Instead, we review recent developments of superconducting radiation sensors based on a structure that shares many technological aspects with the already employed TESs and KIDs. Indeed, the nanoscale transition edge sensor (nano-TES) \cite{Paolucci2} and the Josephson escape sensor (JES) \cite{Paolucci} are envisioned by downsizing the active region of a TES to the nano-scale. In particular, they employ a one dimensional fully superconducting Josephson junction (1D-JJ) as radiation absorber. In full analogy with the widespread TES, these sensors take advantage of the stark variation of the resistance of a superconductor while transitioning to the normal-state. Furthermore, the resistance versus temperature characteristics of a 1D-JJ can be tuned by varying the bias current. As a consequence, in principle the sensitivity of the nano-TES and the JES can be in situ controlled. 
Moreover, the nano-TES and JES intrinsic values of NEP were deduced from the transport and thermal data. The nano-TES is predicted to show a thermal fluctuation limited NEP of about $5 \times 10^{-20}$ W$/\sqrt{\text{Hz}}$ \cite{Paolucci2}, while the JES is expected to provide a NEP of the order of $10^{-25}$ W$/\sqrt{\text{Hz}}$ \cite{Paolucci}. As a consequence, these sensors could be the basis of detectors pointing towards unprecedented sensitivities in the GHz and THz bands.
In addition to gigahertz astronomy, these sensors could be employed in medical imaging \cite{Sun}, industrial quality controls \cite{Ellrich} and security applications \cite{Rogalski}.

This review is organized as follows. Section \ref{Theory} covers the theoretical bases of the nano-TES and JES highlighting their structure and operating principle. Section \ref{TheoryBol} introduces the thermal and electrical model employed to calculate the performance of the sensors, such as noise equivalent power and response time. Section \ref{ExpBol} describes the experimental realization of the nano-TES and the JES. In particular, the nano-fabrication, the thermal, spectral and electronic transport measurements are shown. Section \ref{Perf} reports the sensing performance deduced from the experimental data. Finally, Sec. \ref{Concl} resumes the results and opens to new applications for the nano-TES and the JES detectors.

\section{Sensors structure and operating principle}
\label{Theory}

\begin{figure}[t!]
\centering
\includegraphics[width=\columnwidth]{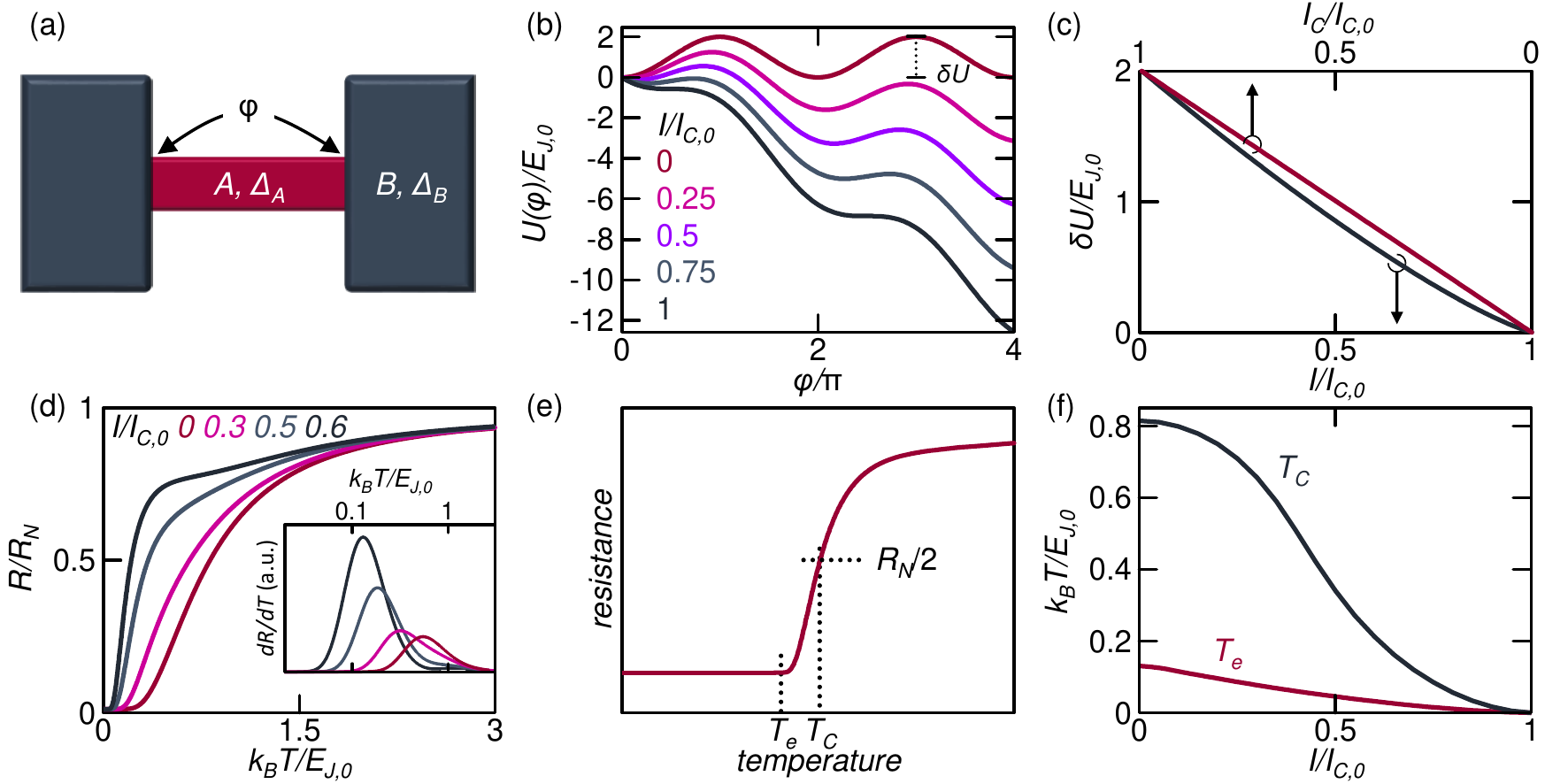}
\caption{Properties of a 1D fully superconducting Josephson junction. (\textbf{a}) Schematic representation of a 1D fully superconducting JJ, where a superconducting nanowrire $A$ is contacted by two superconducting leads $B$ obeying to $\Delta_A\ll \Delta_B$ (with $\Delta_A$ and $\Delta_B$ the superconducting energy gap of the nanowire and the leads, respectively). (\textbf{b}) Energy of the washboard potential normalized with respect to the zero-temperature Josephson energy as a function of the phase difference across a JJ for different values of bias current. The energy  barrier ($\delta U$) for the escape of the phase particle from the WP is indicated. (\textbf{c}) Energy barrier normalized with respect to the zero-temperature Josephson energy ($\delta U/E_{J,0}$) calculated as a function of critical current (top) and bias current (bottom). (\textbf{d}) Resistance (R) versus temperature (T) characteristics calculated for different values of $I$. Inset: Temperature derivative of $R$ calculated for different values of $I$.  (\textbf{e}) Calculated temperature dependence of the JJ resistance, where the escape temperature ($T_e$) and the critical temperature ($T_C$) are indicated. (\textbf{f}) Escape temperature ($T_e$) and critical temperature ($T_C$) calculated as a function of the bias current ($I$).}
\label{FigJJ}
\end{figure} 

The sensors presented in this Review, i.e., the nano-TES \cite{Paolucci2} and the JES \cite{Paolucci}, can be described as a 1D fully superconducting Josephson junction (1D-JJ). A 1D-JJ is realized in the form of a superconducting nanowire ($A$), with both lateral dimensions [thickness ($t$) and width ($w$)] smaller than its coherence length ($\xi_A$) and the London penetration depth of the magnetic field ($\lambda_{L,A}$), separating two superconducting lateral banks ($B$), as schematically shown in Fig. \ref{FigJJ}(a). The 1D nature of the JJ provides a twofold advantage. On the one hand, $\lambda_{L,A}\gg w,t$ ensures homogeneous supercurrent density in the JJ and uniform penetration of $A$ by an out-of-plane magnetic field.
On the other hand, $\xi_A>w,t$ leads to uniform superconducting properties and constant superconducting wave function along the nanowire cross section.

The electronic transport properties of a 1D-JJ can be described through the overdamped resistively shunted junction (RSJ) model \cite{tinkham}, where the dependence of the stochastic phase difference [$\varphi(t)$] over the 1D-JJ on the bias current ($I$) is given by
\begin{equation}
\frac{2e}{\hbar} \frac{\dot{\varphi}(t)}{R_N} + I_C \sin{\varphi(t)} = I + \delta I_{th}(t) \text{,}
\label{CPR}
\end{equation}
where $e$ is the electron charge, $R_N$ is the normal-state resistance, $\hbar$ is the reduced Planck constant and $I_C$ is the critical current. The thermal noise generated by the shunt resistor ($R_N$) is given by $\left\langle \delta I_{th}(t)\delta I_{th}(t')\right\rangle=\frac{K_BT}{R_N}\delta{(t-t')}$, where $k_B$ is the Boltzmann constant and $T$ is the temperature.

Within the RSJ model, the transition of the junction to the resistive state is attributed to $2\pi$ phase-slip events \cite{Barone1982,tinkham}. The phase slips can be viewed as the movements of a phase particle in a tilted washboard potential model (WP) under the action of friction forces. 
In particular, the WP takes the form
\begin{equation}
U(\varphi)=-\frac{\hbar I}{2e}\varphi-\delta U \cos{\varphi}
\text{.}
\label{WP}
\end{equation}
For a 1D-JJ composed of a superconducting nanowire, the escape energy barrier can be written as \cite{Bezryadin2012}
\begin{equation}
\delta U(I,E_J)\sim 2 E_J\left(1-I/I_C \right)^{5/4}=\frac{\Phi_0 I_C}{\pi}\left(1-I/I_C \right)^{5/4}\text{.}
\label{eq:potential}
\end{equation}

Indeed, the effective WP profile strongly depends on both bias current ($I$) through the junction and Josephson energy $E_J=\Phi_0I_C/2\pi$ \cite{Bezryadin2012}, where $\Phi_0\simeq 2.067\times 10^{-15}$ Wb is the flux quantum. In particular, by rising $I$ the phase-slip energy barrier ($\delta U$) lowers, similarly to the result of decreasing $E_J$, and the WP tilts proportionally to the value of the  bias current, as shown in Fig. \ref{FigJJ}(b). Furthermore, Eq. \ref{eq:potential} shows that $\delta U$ is suppressed faster by rising $I$ than by suppressing $I_C$ [see Fig. \ref{FigJJ}(c)]. As a consequence, the transition to the dissipative state can be more efficiently induced by rising the bias current than by decreasing the Josephson energy. 
In addition, we note that $E_J$ (and thus $I_C$) can be suppressed by applying an out-of-plane magnetic field. 
Since a magnetic field can be problematic for several applications and broadens the superconductor-to-dissipative transition \cite{Zant1992}, we will focus on the impact of $I$ on the transport properties of the 1D-JJ.

Since the nanowire normal-state resistance is very low, the 1D-JJ can be efficiently described by the RSJ model in the overdamped limit. Within this model, the resulting voltage ($V$) versus temperature ($T$) characteristics of a 1D-JJ for different values of $I$ and $E_J$ is given by \cite{Ivanchenko1969} 
\begin{equation}
V(I, E_J, T)=R_N \left[ I - I_{C,0} \operatorname{Im}\frac{\mathcal{I}_{1-iz}\left( \frac{E_J}{k_BT}\right) }{\mathcal{I}_{-iz}\left( \frac{E_J}{k_BT}\right)}\right]\text{,}
\label{eq:voltage}
\end{equation}
where $I_{C,0}$ is the zero-temperature critical current and $\mathcal{I}_{\mu}(x)$ is the modified Bessel function with imaginary argument $\mu$ with $z=\frac{E_J}{k_BT}\frac{I}{I_C}$. Thus, the current dependent resistance versus temperature characteristics [$R(T)$] of the 1D-JJ can be evaluated by calculating the current derivative of the voltage drop
\begin{equation}
R(I, E_J, T)=\frac{\text{d}V(I, E_J, T)}{\text{d}I}\text{.}
\label{resistance}
\end{equation}

Figure \ref{FigJJ}(d) shows that $I$ has a twofold effect on the $R(T)$ characteristics. On the one hand, an increase of $I$ lowers the temperature of the superconducting-to-dissipative state transition. On the other hand, the width of the transition narrows by rising $I$. This behavior is also highlighted by the temperature derivative of $R(T)$ [see the inset of Fig. \ref{FigJJ}(d)].

From each $R(T)$ curve we can define two temperatures related to the superconductor-to-normal-state transition: the effective critical temperature ($T_C$) and the escape temperature ($T_e$), as shown in Fig. \ref{FigJJ}(e). In particular, $T_C$ is the temperature corresponding to half of the normal-state resistance [$R(T_C)=R_N/2$], while $T_e(I)$ is the maximum value of temperature providing a zero resistance of the nanowire [$R(T_e)=0$]. These two temperatures define two distinct operating conditions for the 1D-JJ: the nano-TES operates at $T_C$, i.e., at the middle of the superconductor-to-normal-state transition, and the JES operates at $T_e$, i.e., deeply in the superconducting state. Indeed, $T_C$ and $T_e$ strongly depend on the bias current, as shown in Fig. \ref{FigJJ}(f). As a consequence, both nano-TES and JES are \emph{in situ} current-tunable radiation sensors. 

\section{Theory of nano-TES and JES bolometers}
\label{TheoryBol}

\begin{figure}[t!]
\centering
\includegraphics[width=\columnwidth]{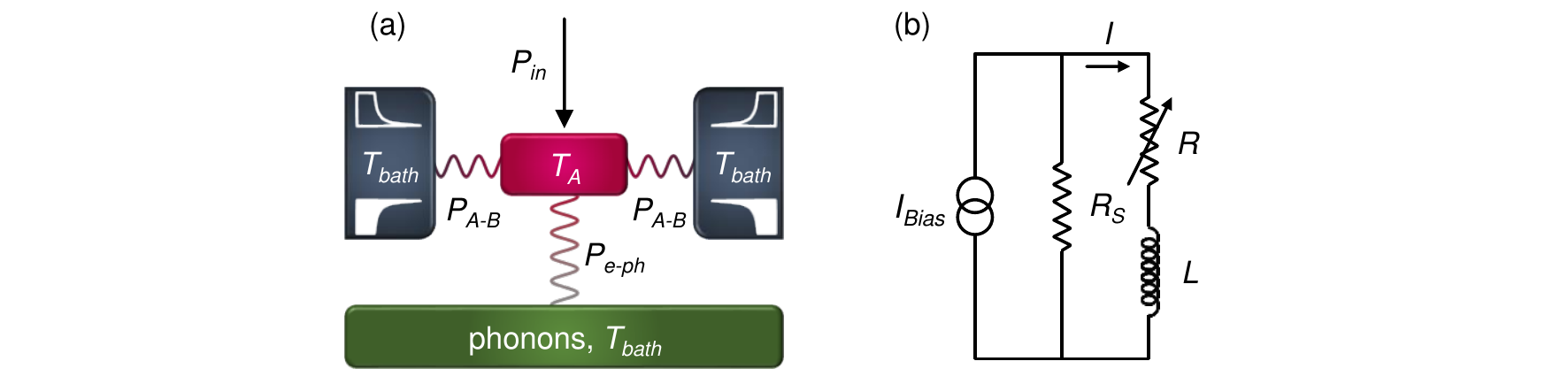}
\caption{General thermal and electrical model of the nano-TES and JES. (\textbf{a}) Thermal model of the nano-sensors where the main thermal exchange channels are shown. 
$P_{in}$ is the power coming from the incoming radiation, $P_{e-ph}$ is the heat exchanged between electrons at $T_A$ (violet) and lattice phonons at $T_{bath}$ (green), and $P_{A-B}$ is the heat current flowing towards the superconducting electrodes at $T_{bath}$ (blue). (\textbf{b}) Schematic representation of a typical biasing circuit for the sensors. The parallel connection of the sensor ($R$) and the shunt resistor ($R_{S}$) is biased by the current $I_{Bias}$. The role of $R_S$ is to limit the Joule overheating of the active region when transitioning to the normal-state. The $I$ variations are measured thank to the inductance $L$, for instance by a SQUID amplifier.}
\label{FigHE}
\end{figure}

The behavior of superconducting bolometers strongly depends on the predominant heat exchange mechanisms occurring in their active region. The  thermal model describing all the main thermal exchange mechanisms occurring in $A$ is the same for the nano-TES and the JES, as schematically shown in Fig. \ref{FigHE}(a). Here, $P_{in}$ is the power associated with the absorption of the external radiation, while $P_{A\text{-}B}$ represents the energy out-diffusion for the active region to the lateral leads kept at the bath temperature ($T_{bath}$). We notice that depending on the device operation, nano-TES or JES, the bath temperature is set to $T_C(I)$  or $T_e(I)$, respectively. The lateral electrodes can act as energy filters ($P_{A\text{-}B}\to 0$), the so-called Andreev mirrors \cite{Andreev1964}, when their critical temperature ($T_{C,B}$) is much higher than the zero-current critical and escape temperatures of the active region [$T_{C,B}\gg T_C(0), T_e(0)$], that are the maximum operating temperatures. Therefore, heat exchange with lattice phonons ($P_{e\text{-}ph}$) is the predominant thermal relaxation channel in the active region. In the TES operation, $A$ is always kept almost in the normal-state thus yielding an electron-phonon coupling \cite{Irwin1995,Giazotto2006}
\begin{equation}
P_{e\text{-}ph,n}=\Sigma_A \mathcal{V}_{A}\left(T_A^5-T_{bath}^5 \right)\text{,}
\label{e-phpowerN}
\end{equation}
where $\mathcal{V}_A$ is the nanowire volume, and $\Sigma_A$ is its electron-phonon coupling constant. On the contrary, since the JES operates at $T_e(I)$, the latter can be substantially smaller than $T_C$ depending on $I$ [see  Fig. \ref{FigJJ}(f)] with $A$ deeply in the superconducting state. The electron-phonon heat exchange in a superconductor takes the form \cite{Timofeev2009}
\begin{equation}
P_{e\text{-}ph,s} \propto P_{e\text{-}ph,n} \exp{[-\Delta_A/(k_B T_A)]}\text{,}
\label{e-phpowerS}
\end{equation}
where $\Delta_A$ is the superconducting energy gap in $A$, thus showing an exponential suppression with respect to the normal-state at very low temperatures. As a consequence, the performance of a JES radiation detector are expected to be considerably better than that of a nano-TES operating at the same temperature.

The transition to the normal-state driven by radiation absorption generates Joule heating in $A$, thus rising $T_A$ and creating thermal instability. For TES bolometers, this issue is solved by a biasing circuit implementing the so-called negative electrothermal feedback (NETF). Similarly, the nano-TES and the JES could be biased with the circuitry shown in Fig \ref{FigHE}(b). The shunt resistor ($R_S$) guarantees that the current ($I$) flowing through the sensor ($R$) is limited when the active region undergoes the superconducting to normal-state transition. In TES operation, the sensor is biased at $T_C$ ($R=R_N/2$) so that the condition for the shunting resistor reads $R_S=IR_N/[2(I_{Bias}-I)]$, where $I_{Bias}$ is the current provided by the generator. In JES-mode, the device has to be biased at $T_e(I)$,i.e., at $R=0$, and the role of $R_S$ is to limit the current flow through the sensing element below the retrapping current ($I_R$) \cite{Courtois}, that is the switching current at which a diffusive superconducting wire switches into the dissipationless state from the normal-state during a current down-sweep. This happens for $R_S\leq R_NI_R/I_{bias}$ and brings $A$ quickly back to the superconducting state after radiation absorption. Therefore, the sensor always operates in the superconducting state. Finally, in both configurations the variations of the current circulating in the inductor ($L$), as a consequence of radiation absorption, can be measured via a conventional SQUID amplifier.

In the next sections, we will show all the theoretical relations describing the sensing properties of the nano-TES and the JES active regions when continuously irradiated (bolometer). Therefore, we will neglect all the noise contributions arising from outside the sensors, such as the noise coming from the electronic circuitry \cite{Giazotto2006}.

\subsection{Nano-TES bolometer performance}
\label{TheoTES}
The noise equivalent power (NEP) represents the minimum power that can be detected above the noise level. 
Thus, the NEP is the most important figure of merit for a bolometer based on the nano-TES sensing element.
Taking into account the equivalent circuit shown in Fig. \ref{FigHE}, the total $NEP$ of the nano-TES is given by the \cite{Mather,Lee} 
\begin{equation}
    NEP_{tot} = \sqrt{NEP^2_{TFN,nano\text{-}TES} + NEP^2_{Jo} + NEP^2_{R_S}}\text{,}
    \label{Eq:totalNEP}
\end{equation}
where ${NEP_{TFN,nano\text{-}TES}}$ is associated with the thermal fluctuations, ${NEP_{Jo}}$ is due to the Johnson noise, while ${NEP_{R_S}}$ arises from the shunt resistor.

For a nano-TES operating at $T_A\simeq T_C$, the thermal fluctuations limited $NEP$ takes the form \cite{Bergmann}\\
\begin{equation}
NEP_{TFN,nano\text{-}TES} = \sqrt{4\Upsilon G_{th,nano\text{-}TES} k_{B} {T_C}^{2}}\text{,}
\label{NEP_TES}
\end{equation}
where $\Upsilon = n /(2n+1)$ accounts for the temperature gradient between the quasiparticles in $A$ and all the other thermal sinks (with $n = 5$ for pure and $n=4,6$ for dirty metals), and ${G_{th,nano\text{-}TES}}$ is the thermal conductance of all heat losses.\\
Since $T_C\ll T_{C,B}$, the main thermalization channel for $A$ is the electron-phonon coupling ($P_{e\text{-}ph}$). Furthermore, the 1D-JJ is always kept in the partially dissipative state (at $R_N/2$) for the nano-TES operation. So, the electron-phonon coupling of a normal-metal diffusive thin film ($P_{e\text{-}ph, n}$, see Eq. \ref{e-phpowerN}) is employed \cite{Giazotto2006}. The resulting thermal conductance for a nano-TES ($G_{th,nano\text{-}TES}$) can be calculated through the temperature derivative of the electron-phonon energy relaxation \cite{Irwin1995}
\begin{equation}
G_{th,nano\text{-}TES} = \dfrac{\text{d} P_{e\text{-}ph, n} }{\text{d} T_A} = 5 \Sigma_A {\mathcal{V}}_{A} T_A^4\text{.}
\label{ThermalConductance}
\end{equation}

The Johnson noise is originated by the electronic transport through $A$ when the nano-TES is in the normal-state  and takes the form \cite{Bergmann}
\begin{equation}
    NEP_{Jo}  = \sqrt{2 k_B R_N T_{C}} \frac{G_{th,nano\text{-}TES} T_{C}}{V \alpha} \sqrt{1+ 4\pi^2 f^2 \tau_{eff}^2} \text{,}
    \label{Eq:NEP_Jo}
\end{equation}
where $V$ is the voltage drop, $f$ is the signal bandwidth, $\alpha = \dfrac{\text{d} R}{\text{d} T}\dfrac{T}{R}$ is the electrothermal parameter accounting for the sharpness of the phase transition from the superconducting to the normal-state \cite{Irwin1995}, and $\tau_{eff}$ is the effective circuit time constant (see Eq. \ref{taueff}).

Finally, the charge fluctuations through $R_{S}$ provide a contribution \cite{Bergmann}
\begin{equation}
            NEP_{R_S}  = \sqrt{4 k_B R_{S}  T_{bath}} \frac{G_{th,nano\text{-}TES} T_{C}}{V \alpha}\sqrt{(1-L_0)^2 + 4\pi^2 f^2 \tau_{eff}^2}, 
       \label{Eq:NEP_Sh}
\end{equation}    
where $L_0 = \alpha/n$ is the loop gain. Since the shunting resistor needs to satisfy $R_{sh}\ll R_A$, the contribution of $NEP_{sh}$ is usually negligible compared to Johnson noise. Indeed, in superconducting bolometers, the thermal fluctuations contribution dominates over ${NEP_{Jo}}$ and ${NEP_{R_S}}$. Thus, we can consider $NEP_{tot}=NEP_{TFN,nano\text{-}TES}$.

The time response of the detector can be calculated by solving the time dependent energy balance equation that takes into account all the exchange mechanisms after radiation absorption \cite{Giazotto2006}. In particular, the re-thermalization (to $T_{bath}$) of the quasiparticles in $A$ has an exponential dependence on time. The associated time constant ($\tau_{nano\text{-}TES}$) can be calculated as the ratio between the thermal capacitance and the thermal conductance of $A$
\begin{equation}
\tau_{nano\text{-}TES} = \dfrac{C_{e,nano\text{-}TES}}{G_{th,nano\text{-}TES}}\text{.}
\label{IntrinsicTime}
\end{equation}
where $C_{e,nano\text{-}TES}$ is the electron heat capacitance. The latter is written\\
\begin{equation}
C_{e,nano\text{-}TES} = \gamma_A \mathcal{V}_{A} T_C\text{,}
\label{HeatCapacitance}
\end{equation}
where $\gamma_A$ is the Sommerfeld coefficient of the active region. 

We note that $\tau_{nano\text{-}TES}$ is the intrinsic recovery time of $A$. This term does not take into account the Joule heating due to the current flowing through the sensor in the dissipative state. By considering the circuitry implementing the NETF [see Fig. \ref{FigHE}(b)], the pulse recovery time becomes \cite{Irwin1995}\\
\begin{equation}
\tau_{eff} = \dfrac{\tau_{nano\text{-}TES}}{1 + \dfrac{\alpha}{n}}\text{.}
\label{taueff}
\end{equation}

For $\tau_{eff}<< \tau_{nano\text{-}TES}$, i.e., when the pulse recovery time is much shorter than the intrinsic time constant of $A$, the overheating into the active region is decreased, thus compensating for the initial temperature variation and avoiding the dissipation through the substrate.

\subsection{JES bolometer performance}
\label{TheoJES}

The JES operates at $T_e(I)$, that is fully in the superconducting state. Since the current injection does not change the energy gap of the active region ($\Delta_A\sim \text{const}$), only the effective critical temperature of $A$ changes with $I$, while the intrinsic values of critical temperature ($T_C^i$) is unaffected. As a consequence, all figures of merit of a JES are calculated deeply in the superconducting state, thus ensuring high sensitivity.

In full analogy with the nano-TES, the thermal fluctuations are the limiting factor for the sensitivity of a JES bolometer. The related contribution to the NEP reads
\begin{equation}
{NEP}_{TFN,JES} = \sqrt{4\Upsilon {G_{th,JES}} k_{B} {T_e}^{2}}\text{,}
\label{NEP_JE}
\end{equation}
where $G_{th,JES}$ is the thermal conductance in the superconducting state. The latter takes the form \cite{Heikkila}

\begin{equation}
G_{th,JES} \approx \dfrac{\Sigma_A \mathcal{V}_A {T_e}^4}{96 \varsigma(5)}
\left[ f_1 \left(\dfrac{1}{\tilde{\Delta}}\right) \cosh(\tilde{h}) e^{-\tilde{\Delta}} + \pi \tilde{\Delta}^5 f_2 (\dfrac{1}{\tilde\Delta})e^{-2 \tilde{\Delta}}  \right] \text{,}
\label{eqGJES}
\end{equation}
where the first term refers to the electron-phonon scattering, while the second term stems from the recombination processes. In Eq. \ref{eqGJES}, $\varsigma(5)$ is the Riemann zeta function, $\tilde{\Delta} = \Delta_A/k_B T$ is the normalized energy gap of $A$, $\tilde{h} = h/k_B T$ represents exchange field (0 in this case), $f_1(x) = \begin{matrix} \sum_{n=0}^3  C_n x^n \end{matrix} $ with $C_0 \approx 440, C_1 \approx 500, C_2 \approx 1400, C_3 \approx 4700$, and $f_2(x) = \begin{matrix} \sum_{n=0}^2  B_n x^n \end{matrix}$ with $B_0 = 64, B_1 = 144, B_2= 258$. We note that the thermal conductance for a JES is exponentially damped compared to the nano-TES, due to the operation in the superconducting state. Thus, we expect the JES to be extremely more sensitive than the nano-TES, that is ${NEP}_{TFN,JES}\ll {NEP}_{TFN,nano\text{-}TES}$.

The JES speed in given by the relaxation half-time ($\tau_{1/2}$), which reads \cite{Virtanen}\\
\begin{equation}
\tau_{1/2} = \tau_{JES} \ln{2} \text{,}
\label{TauMezzi} 
\end{equation}
where $\tau_{JES}$ is the JES intrinsic thermal time constant. The latter is calculated by substituting the JES parameters in Eq. \ref{IntrinsicTime}, thus considering ${C_{e,JES}}$ and ${G_{th,JES}}$ in deep superconducting operation. The electron heat capacitance needs to be calculated at the current-dependent escape temperature [$T_e(I)$], thus in the superconducting state, and takes the form\\
\begin{equation}
{C_{e,JES}} = \gamma_{A} \mathcal{V}_{A}T_e \Theta_{Damp}={C_{e,nano\text{-}TES}}\Theta_{Damp}\text{,}
\end{equation}
where $\Theta_{Damp}$ is the low temperature exponential suppression with respect to the normal metal value. The suppression is written \cite{Rabani}\\
\begin{equation}
\Theta_{Damp} = \dfrac{C_s}{1.34\gamma_{A} T_e}\text{.}
\end{equation}

Finally, the electronic heat capacitance is given by
\begin{equation}
C_s = 1.34 \gamma_{A} T_C^i \left( \dfrac{\Delta_A}{k_B T_e} \right)^{-3/2} e^{\Delta_A/k_B T_e}\text{.}
\end{equation}

\section{Experimental realization of JES and nano-TES}
\label{ExpBol}

The  nano-TES and JES are experimentally realized and tested thanks to two different device architectures: an auxiliary device and the proper nano-sensor. On the one hand, the auxiliary device was employed to determine the superconducting energy gap and the thermal properties of the material composing the one-dimensional active region. On the other hand, the measurements performed on the proper nanosensor provided dependence of the $R(T)$ characteristics on the bias current.

\subsection{Fabrication procedure}
Both devices were fabricated during the same evaporation process, ensuring the homogeneity of the properties of $A$. In particular, they were fabricated by electron-beam lithography (EBL) and 3-angles resist shadow mask evaporation onto a silicon wafer covered with 300 nm of thermally grown SiO$_{2}$. 
To obtain a resist suspended mask, a bilayer composed of a 950-nm-thick MMA(8.5)MMA layer and a PMMA (A4, 950k) film of thickness of about 300 nm was employed. The ratio between the electron irradiation doses to make the resists soluble is about 1 to 4. 
The evaporation was performed in an ultra-high vacuum electron-beam evaporator with a base pressure of about $10^{-11}$ Torr by keeping the target substrate at room temperature. The first step was the evaporation of a 13-nm-thick Al layer at an angle of -40$^\circ$ and then oxidized by exposition to 200 mTorr of O$_{2}$ for 5 minutes to obtain the tunnel probes of the auxiliary device. In a second step, the Al/Cu bilayer ($t_{Al} = 10.5$ nm and $t_{Cu} = 15$ nm) forming the active region is evaporated at an angle of $0^\circ$. Finally, a second 40-nm-thick Al film was evaporated at an angle of $+40^\circ$ to obtain the lateral electrodes. The angle resolution of the evaporation was about 1$^\circ$.
The average film thickness can be controlled with the precision of 0.1 nm at the evaporation rate of about 1.5 angstrom/s. This evaporation rate allows good thickness control and films with state-of-the-art diffusion constants and superconducting properties. Finally, the tunnel barriers present in the auxiliary device were reproducible and almost free from pin-holes.

\subsection{Density of states and thermal properties of the active region}
\label{DOSTHE}

\begin{figure}[t!]
\centering
\includegraphics[width=\columnwidth]{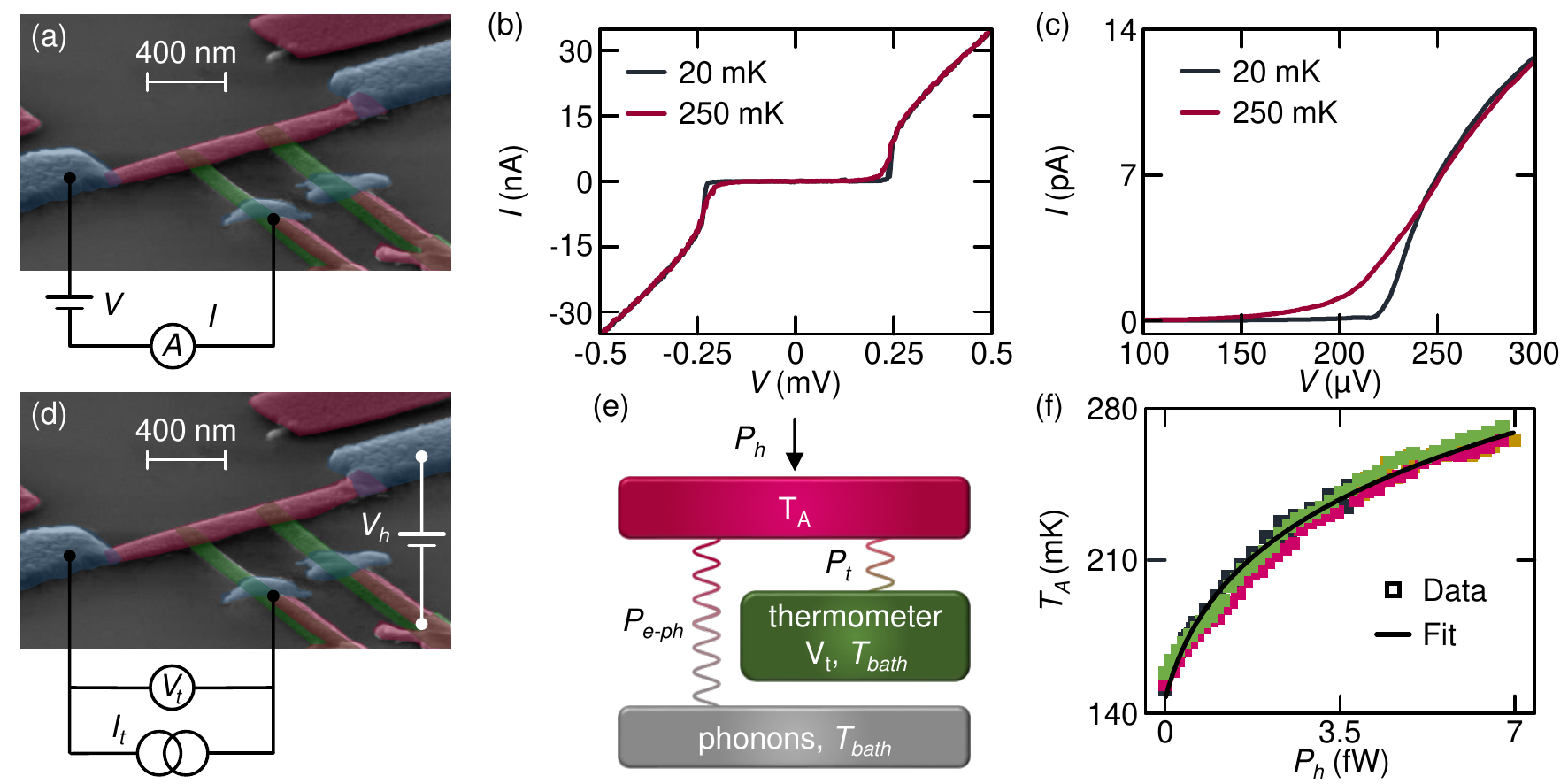}
\caption{Energy gap and thermal properties of the nano-sensor active region. (\textbf{a}) False-color scanning electron micrograph of a typical auxsiliary device. The core of the device is made of an Al/Cu bilayer nanowire (purple) sandwiched between two Al leads (blue). To perform tunnel spectroscopy a voltage ($V$) is applied between $A$ and an Al tunnel probe (green) while recording the current ($I$). (\textbf{b}) Tunneling current ($I$) versus voltage ($V$) characteristics recorded at $T_{bath}=20$ mK (blue) and $T_{bath}=250$ mK (purple). (\textbf{c}) Zoom of the $IV$ characteristics in correspondence of the transition to the normal-state. It is possible to extract $\Delta_{A,0}\simeq23\;\mu$eV and $\Delta_{P,0}\simeq200\;\mu$eV. (\textbf{d}) Schematic representation of the set up employed for the thermal measurements. The left tunnel junction serving as thermometer is biased with a constant current ($I_t$) while the voltage ($V_t$) is measured. The right junction is voltage biased ($V_h$) and operated as local electronic heater. (\textbf{e}) Thermal model describing the main heat exchange mechanisms involved in the experiment. The power injected by the local heater ($P_h$) relaxes by the exchange with phonons ($P_{e\text{-}ph}$) and outflows through the thermometer tunnel junction ($P_t$). (\textbf{f}) Electronic temperature of the active region ($T_A$) versus the injected power by the heater ($P_h$). The dots represent 5 different sets of measurements, while the back line depicts the fit performed by employing Eq. \ref{balance_equation}.}
\label{FigGT}
\end{figure}

The false-color scanning electron microscope (SEM) picture of a typical auxiliary device is shown in Fig. \ref{FigGT}(a). The 1D-JJ is formed by 1.5 $\mu$m-long ($l$), 100 nm-wide ($w$) and 25 nm-thick ($t$) Al/Cu bilayer nanowire-like active region (purple) sandwiched between the two Al electrodes (blue). To characterize both the energy gap and the thermal properties of $A$, the device is equipped with two additional Al tunnel probes (green).

Figure \ref{FigGT}(a) schematically represents the experimental set up employed to carry out the spectral characterization of the active region in a filtered He$^3$-He$^4$ dry dilution refrigerator. The $IV$ tunnel characteristics of $A$ are performed by applying a voltage ($V$) and measuring the current ($I$) flowing between one lateral electrode and a tunnel probe.

For $T<0.4T_C$, the energy gap of a superconductor follows $\Delta(T)=\Delta_0$, where $\Delta_0$ is its zero-temperature value \cite{tinkham}. Typically, aluminum thin films show a $T_C\geq 1.2$ K, that is the bulk Al critical temperature \cite{cochran}. Therefore, the superconducting gap of the superconducting tunnel probes is temperature independent up to at least 500 mK. On the contrary, superconductivity in $A$ is expected to be strongly suppressed due to inverse proximity effect \cite{tinkham}. Therefore, this structure allows to determine the zero-temperature superconducting energy gap of the active region ($\Delta_{0,A}$). 
The $IV$ characteristics were measured at the base temperature ($T=20$ mK) and well above the expected critical temperature of $A$, as shown in Fig. \ref{FigGT}(b). At the base temperature, the JJ switches to the normal-state when the voltage bias reaches $V=\pm(\Delta_{A,0}+\Delta_{P,0})/e$ \cite{Giazotto2006}, where $\Delta_{0,P}$ is the zero-temperature gap the Al probe. Instead, at $T_{bath}= 250$ mK the transition occurs at $V=\pm\Delta_{P,0}/e$, since $A$ is in the normal-state. In addition, the tunnel resistance of the JJ takes the values $R_I\simeq 12$ k$\Omega$. The difference between the curves recorded at 20 mK and and 250 mK is highlighted by Fig. \ref{FigGT}(c). The measurement at $T_{bath}= 250$ mK (purple) shows a value of the zero-temperature energy gap of the Al probe $\Delta_{0,P}\simeq200\;\mu$eV, indicating a critical temperature $T_{C,P}=\Delta_{P,0}/(1.764k_B)\simeq 1.3$ K.
Furthermore, the difference between the results obtained at 20 mK and 250 mK leads to $\Delta_{A,0}\simeq23\;\mu$eV pointing towards a critical temperature $T_{C,A}\simeq150$ mK.

We now focus on the heat exchange properties of the active region. The schematic representation of the experimental setup employed is shown in Fig. \ref{FigGT}(d). The left Josephson junction operates as a thermometer: it was current-biased at $I_{t}$ and the resulting voltage drop ($V_t$) reflects the variations of the electronic temperature in $A$ \cite{Giazotto2006}. Instead, the right JJ was voltage-biased at $V_{h}>\Delta_{A,0}+\Delta_{P,0})/e$ to work as heater \cite{Giazotto2006}. 

The geometry of the device guarantees that the electronic temperature of the lateral superconducting electrodes and the tunnel probes are equal to the phonon bath temperature ($T_{bath}$). By contrast, the quasiparticles temperature of the active region is the fundamental thermal variable in the experiment.
Since $T_A\ll T_{C,B}$ for all the measurements, the lateral banks serve as perfect Andreev mirrors ($P_{A\text{-}B}= 0$). Therefore, the power injected by the heater ($P_{h}$) relaxes only via electron-phonon interaction $P_{e-ph}$ and out-diffuses through the tunnel junction acting as thermometer ($P_{t}$). The resulting quasi-equilibrium thermal exchange model is
\begin{equation}
P_{h}= P_{e\text{-}ph}+P_{t}\text{,}
\label{balance_equation}
\end{equation}
as schematically represented in Fig. \ref{FigGT}(e). Since the active region is in the normal-state, the power exchanged via electron-phonon interaction is described by Eq. \ref{e-phpowerN}, while the power flowing through the thermometer reads \cite{Giazotto2006}
\begin{equation} 
P_{t}=\frac{1}{e^{2} R_{t}} \int_{-\infty}^{+\infty} \mathrm{d}E  E\;DOS_{P}(E,T_{bath})\left[ f_{0}(E-eV_t,T_{A})-f_{0}(E,T_{bath})\right],
\label{NISpower}
\end{equation}
where $R_t=11.6$ k$\Omega$ is the normal-state tunnel resistance of the thermometer, and  $f_{0}(E,T_{A,bath})=\left[1+\exp{\left(E/k_{B}T_{A,bath}\right)}\right]^{-1}$ is the Fermi-Dirac distribution of the active region and the superconducting probe, respectively. Above, the normalized density of states of the superconducting probe takes the form \cite{Giazotto2006}:
\begin{equation}
DOS_P(E, T_{bath})=\frac{|E|}{\sqrt{E^2-\Delta_P^2(T_{bath})}}\Theta(E^2-\Delta_P^2(T_{bath})).
\label{eq_dos}
\end{equation}

Figure \ref{FigGT}(f) shows the dependence of electronic temperature in $A$ on the injected power ($P_h$) acquired at $T_{bath}=150$ mK for five different sets of measurements. The value of $T_A$ rises monotonically from $T_{bath}=150$ mK to about $270$ mK by increasing $P_{h}$ up to $\sim7$ fW.

By solving Eq. \ref{balance_equation}, it is possible to fit the experimental values of $T_A$ as a function of $P_{h}$. 
This simple model is in good agreement with the data, as shown by the black line in Fig. \ref{FigGT}(f). The extracted value of the electron-phonon coupling constant of the Al/Cu bilayer is $\Sigma_{A}\simeq 1.3\times 10^{9}$ W/m$^{3}$K$^{5}$ in good agreement with the average of $\Sigma_{Cu}=2.0\times10^{9}$ W/m$^{3}$K$^{5}$ and $\Sigma_{Al}=0.2\times10^{9}$ W/m$^{3}$K$^{5}$ \cite{Giazotto2006}, weighted with the volumes of the copper and the aluminum layer forming the active region: $\Sigma_{A,theo}=(\Sigma_{Cu}\mathcal{V}_{Cu}+\Sigma_{Al}\mathcal{V}_{Al})/\mathcal{V}_{A}=1.38\times10^9$ W/m$^{3}$K$^{5}$, with $\mathcal{V}_{Al}\simeq1.58\times 10^{-21}$ m$^{-3}$, $\mathcal{V}_{Cu}\simeq2.25\times 10^{-21}$ m$^{-3}$, and $\mathcal{V}_{A}=\mathcal{V}_{Al}+\mathcal{V}_{Cu}\simeq3.83\times 10^{-21}$  m$^{-3}$ the resulting total volume of $A$. 

Since the fit provides $P_{e\text{-}ph}\gg P_t$, the presence of the thermometer tunnel barrier has a negligible impact on the determination of $\Sigma_A$. Furthermore, the electronic temperature in $A$ varies for distances of the order of the electron-phonon coherence length $l_{e-ph}=\sqrt{D_{A}\tau_{e-ph}}\simeq180\;\mu$m \cite{Giazotto2006}, where $D_{A}=(t_{Al}D_{Al}+t_{Cu}D_{Cu})/(t_{Al}+t_{Cu})\simeq5.6\times10^{-3}\text{m}^2/\text{s}$ is the diffusion constant of the active region and $\tau_{e-ph}=C/G\simeq6\;\mu$s is the electron-phonon scattering time. Since the length of the active region is $l=1.5\;\mu$m$ \ll l_{e-ph}$, the electronic temperature in $A$ can be assumed homogeneous.

\subsection{Bias current control of resistance versus temperature characteristics}

\begin{figure}[t!]
\centering
\includegraphics[width=\columnwidth]{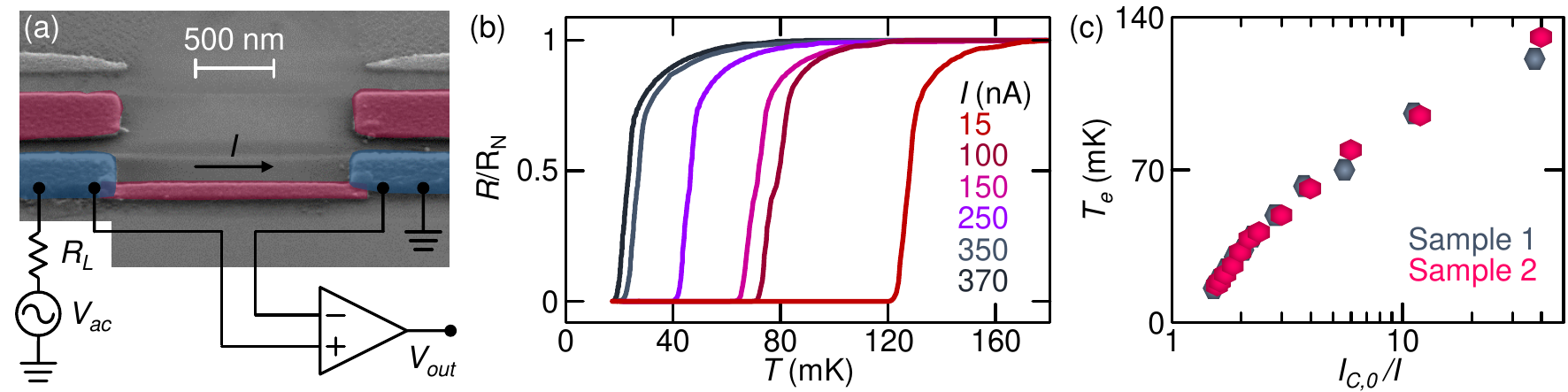}
\caption{Experimental $I$-tuning of the transport properties of a 1D-JJ. (\textbf{a}) False-color scanning electron micrograph of a typical 1D-JJ. The core of the device is made of an Al/Cu nanowire (purple) in contact with thick Al leads (blue).
The nanosensor is AC current-biased (amplitude $I$), while the voltage drop across the wire ($V_{out}$) is measured through a lock-in amplifier. $R_L$ is a load resistor ($R_L \gg R_N$). (\textbf{b}) Selected normalized resistance ($R/R_N$) versus temperature ($T$) characteristics recorded for different values of bias current ($I$). (\textbf{c}) Dependence of the escape temperature ($T_e$) on the ratio between the zero-temperature critical current and the bias current ($I_{C,0}/I$) for two different 1D-JJs.}
\label{FigCC}
\end{figure}

The realization of a typical 1D-JJ is shown by the scanning electron micrograph displayed in Fig. \ref{FigCC}(a).
The 1D superconducting active region consists of the same Al/Cu bilayer presented in Sec. \ref{DOSTHE}, that is $t_{Al}=10.5$ nm and $t_{Cu}=15$ nm. The width of $A$ is $w=100$ nm, while its length is $l=1.5\;\mu$m.  Finally, the lateral banks are composed of a 40-nm-thick aluminum film. The zero-temperature critical current of this 1D-JJ is $I_{C,0}\simeq575$ nA.

The resistance $R$ vs temperature characteristics and of the Al banks were obtained by conventional four-wire low-frequency lock-in technique at $13.33$ Hz in a filtered He$^3$-He$^4$ dry dilution refrigerator. The current was generated by applying a voltage ($V_{ac}$) to a load resistor ($R_L$) of impedance larger than the device resistance ($R_L=100$ k$\Omega\gg R$), as shown in Fig. \ref{FigCC}(a). In particular, the device showed a normal state resistance $R_N\simeq 77\;\Omega$.

To prove that the Al/Cu bilayer can be considered a uniform superconductor, we need to demonstrate that it respects the Cooper limit \cite{DeGennes1964,Kogan1982}, that is the contact resistance between the two layers is negligible and the thickness of each component is lower than its coherence length. Since its large surface area, the Al/Cu interface can be considered fully transparent, i.e., its resistance is negligibly small with comparison to $R_N$. Furthermore, the superconducting Al film fulfils $\xi_{Al}=\sqrt{\hbar D_{Al}/\Delta_{Al}}\simeq80$ nm $\gg t_{Al}=10.5\;\text{nm}$, where $D_{Al}=2.25\times 10^{-3}$ m$^2$s$^{-1}$ is  the diffusion constant of Al thin films and $\Delta_{Al}\simeq200\;\mu$eV is its superconducting energy gap. Concurrently, the normal copper film respects $\xi_{Cu}=\sqrt{\hbar D_{Cu}/(2\pi k_B T)} \simeq255$ nm $\gg t_{Cu}=15\;\text{nm}$, with $D_{Cu}=8\times 10^{-3}$ m$^2$s$^{-1}$ the typical copper diffusion constant for thin films deposited in the same conditions (pressure and evaporation rate. We note that the temperature is chosen to the worst case scenario $T=150$ mK. As a consequence, the Al/Cu bilayer lies within the Cooper limit ad can be considered a single superconducting material.

We now focus on the 1D nature of the fully superconducting nanowire Josephson junction. The superconducting coherence length in $A$ is given by $\xi_A=\sqrt{l \hbar/[(t_{Al}N_{Al}+t_{Cu}N_{Cu}) R_Ne^2\Delta_{A,0}]}\simeq 220$ nm, where $N_{Al}=2.15\times 10^{47}$ J$^{-1}$m$^{-3}$ and $N_{Cu}=1.56\times 10^{47}$ J$^{-1}$m$^{-3}$ are the density of states at the Fermi level of Al and Cu, respectively. Therefore, the bilayer shows a constant pairing potential along the out-of-plane axis, since the Cooper pairs coherence length is much larger than its thickness ($\xi_a\gg t=t_{Al}+t_{Cu}=25.5\;\text{nm}$). Furthermore, $\xi_A\gg w=100$ nm. So, the active region is one dimensional with respect to the superconducting coherence length. In addition, the London penetration depth for the magnetic field of $A$ can be calculated from $\lambda_{L,A}=\sqrt{(\hbar wt_{A} R_N)/(\pi \mu_0 l \Delta_{A,0})}\simeq 970$ nm, where $\mu_0$ is the magnetic permeability of vacuum. Since $\lambda_{L,A}\gg t,w$, $A$ is 1D with respect to the London penetration depth, too.

The magnetic field generated at the wire surface by the maximum possible bias current is $B_{I,max}=\mu_0 I_{C,0}/(2\pi t)\simeq 5$ $\mu$T. This value is orders of magnitude lower than the critical magnetic field of $A$ that was measured to be about 21 mT \cite{Paolucci}. So, the self generated magnetic field does not affect the properties of the device. Furthermore, the superconducting properties of the Al/Cu bilayer dominate the behavior of $A$. In fact, the energy gap expected for a non-superconducting Al/Cu bilayer as originated by lateral proximity effect is $E_g\simeq3\hbar D_A/l^2\simeq5\;\mu$eV \cite{Golubov}. This value is about 1/4 of the measured $\Delta_A$ (see Sec. \ref{DOSTHE}).

The device in Fig.\ref{FigCC}(a) fulfills all the requirements of a 1D-JJ (see Sec. \ref{Theory}). Therefore, it can be used to investigate the impact of $I$ on the $R(T)$ characteristics. To this end, excitation currents with amplitude ranging from 15 nA to 370 nA were imposed through the device. Figure \ref{FigCC}(b), shows that the $R(T)$ characteristics monotonically move towards low temperatures by rising the current from $\sim 3\%$ and $\sim 65\%$ of $I_{C,0}$. Furthermore, the resistance versus temperature characteristics preserve the same shape up to the largest bias currents. We stress that the use of an AC bias allowed to resolve the $R$ vs $T$ characteristics near the critical temperature. In fact, values of DC bias higher than the retrapping current ($I_{DC}>I_R$) would cause the sudden transition of the device resistance to $R_N$. Instead, the AC bias has always a part of the period lower than $I_R$ thus enabling the precise measurement of the entire $R(T)$ traces. In order to maximize the part of the period with $I\sim I_C$, the AC signal should be a square wave with the upper level close to $I_C$ and the lower value obeying to $I_L < I_R$. If the duty cycle of the signal is 95 $\%$ in the upper level, the sensor operates most of the time near $I_C$ thus providing high sensitivity.

Despite the $R(T)$ curves shift towards low temperatures by increasing the bias current, the active region electronic temperature ($T_A$) at the middle of the superconducting-to-dissipative-state transition under current injection does not coincide with $T_{bath}$. In fact, Joule dissipation (for $R\ne0$) causes the quasiparticles overheating in $A$ yielding $T_A > T_{bath}$. Therefore, the operation of the sensor as a nano-TES is not possible without the additional shunting resistor [see Fig. \ref{FigHE}(b)]. By contrast, the operation as a JES (at $R=0$) guarantees that the electronic temperature of $A$ coincides with the bath temperature.

From the $R$ vs $T$ curves we can specify the current-dependent escape temperature [$T_e(I)$]. The values of $T_e$ are shown in Fig. \ref{FigCC}(c) as a function of $I_{C,0}/I$ for two different devices. The escape temperature is monotonically reduced by rising the bias current with a minimum value $\sim20$ mK for $I=370$ nA, that is $\sim 15\%$ of the intrinsic critical temperature of the active region, $T_C^i\sim 130$ mK. As a consequence, the bias current is the ideal tool to in situ tune the properties of the 1D-JJ when operated as a JES.

\section{Performance deduced from the experimental data}
\label{Perf}
Here, we show the deduced performance of the 1D-JJ based bolometers operated both as nano-TES (at $T_C$) and JES (at $T_e$). To this end, the experimental data ($\Delta$, thermal and transport properties of $A$) showed in Sec. \ref{ExpBol} are substituted in the theoretical models presented in Sec. \ref{TheoryBol}.

\subsection{Nano-TES bolometer experimental deduced performance}

\begin{table}
\caption{Performance of nano-TES bolometers deduced from the experimental data reported in Sec. \ref{ExpBol}.}
\label{TabTES}
\centering
\begin{tabular}{cccccc}
\toprule
\textbf{nano-TES}	& \textbf{$T_C$} (mK) & \textbf{$\alpha$}	& \textbf{$NEP_{TFN, nano\text{-}TES}$} (W/$\sqrt{\text{Hz}}$) & \textbf{$\tau$} ($\mu$s) & \textbf{$\tau_{eff}$} (ns) \\
\midrule
Sample 1	& 128		&2742	& $5.2\times 10^{-20}$  &6      &10\\
Sample 2	& 139		&121	& $6.7\times 10^{-20}$  &5      &200\\
\bottomrule
\end{tabular}
\end{table}

The thermal fluctuations between the electrons and phonons in the active region are the limiting factor for the \textcolor{blue}{intrinsic} noise equivalent power of a nano-TES \cite{Giazotto2006}, as described in Sec. \ref{TheoTES}. Indeed, the contributions related to the Johnson noise and the shunting resistor are negligibly small \cite{Paolucci2}. Despite the $R(T)$ characteristics are modulated by the bias current, Joule heating restricts the operation of the nano-TES at $I\to 0$. As a consequence, the values of $NEP_{TFN,nano\text{-}TES}$ can be only extracted for the lowest experimental values of $I$. 

Table \ref{TabTES} shows the $NEP_{TFN,nano\text{-}TES}$ of two different 1D-JJs. In particular, the thermal fluctuations limited $NEP$ of sample 1 ($NEP_{TFN,1}\simeq 5.2\times 10^{-20}$ W/$\sqrt{\text{Hz}}$) is lower than that of sample 2, since it can be operated at lower temperature ($T_{C,1}<T_{C,2}$) and its thermal losses are reduced ($G_{th,1}\simeq 6.7\times 10^{-15}$ W/K and $G_{th,2}\simeq 9.3\times 10^{-15}$ W/K). This sensitivity is one order of magnitude better than state-of-the-art transition edge sensors \cite{Khosropanah}, since $G_{th, nano\text{-}TES}$ is drastically reduced thanks to the small dimensions of $A$ and the presence of efficient Andreev mirrors.

The saturation power of the nano-TES can be written as
\begin{equation}
P_{sat} = \left(1- \frac{R(T_C)}{R_N}\right)P_{e-ph},
\end{equation}
where $R(T_C)$ $\simeq40$ $\Omega$ is the resistance of the active region at $T_C$. The saturation power is of the order of tens of hundreds of aW. It is possible to increase the saturation power by increasing the heat losses through the phonons, that is by increasing the active region volume, with the simultaneous increase of the $NEP$.

The intrinsic time constant of the 1D-JJ ($\tau$) can be calculated from Eq. \ref{IntrinsicTime}. Here, the heat capacitance is given by Eq. \ref{HeatCapacitance} taking the value $C_{e,1}=\simeq 4\times 10^{-20}$ J/K and $C_{e,2}=\simeq 4.2\times 10^{-20}$ J/K for sample 1 and 2, respectively, where the Sommerfeld coefficient of the active region is calculated as the average of the two components $\gamma_{A}=(\gamma_{Cu}\mathcal{V}_{Cu}+\gamma_{Al}\mathcal{V}_{Al})/\mathcal{V}_{A}$ (with $\gamma_{Cu}=70.5$ JK$^{-2}$m$^{-3}$, $\gamma{Cu}=91$ JK$^{-2}$m$^{-3}$). 

The presence of the bias circuit producing the NETF speeds up the response of a nano-TES bolometer. In fact, effective detector time constant becomes $\tau_1\simeq10$ ns for sample 1 and $\tau_2\simeq200$ ns for sample 2. The difference between the two devices arises from the values of the electrothermal parameter ($\alpha_1\simeq2742$ and $\alpha_2\simeq121$, respectively), as shown by Eq. \ref{taueff}. 

It is interesting to note that, in principle, the nano-TES is fully tunable through the current injection. In fact, the critical temperature, i.e., the temperature giving $R(T_C)=R_N/2$, can be varied by $I$, as shown in Fig. \ref{FigJJ}(d). In order to be $I$-tunable, the electronic temperature of the active region should be not affected by Joule overheating. This condition is fulfilled when $I^2R_N\ll P_{e\text{-}ph,n}$, that is when electron-phonon relaxation is able to fully balance the Joule dissipation in $A$.

\subsection{JES bolometer experimental deduced performance}

Also for the JES, the noise equivalent power is limited by thermal fluctuations between the electron and phonon systems in $A$ \cite{Giazotto2006}. Other limitations to the resolution can arise from the measurement of the superconducting to normal-state transition, which we assume it is optimized to be sub-dominant.
The values of $NEP_{TFN,JES}$ can be extracted by substituting in Eq. \ref{NEP_JE} the measured parameters of the JES, such as $I_{C,0}$, $R(T)$ characteristics, $\Delta_A$ and $\Sigma_A$ (see previous sections for details). We note that the performance can be extracted at any value of $I$, since the operation in the dissipationless superconducting state ensures that $T_e(I)=T_{bath}$ [see Fig. \ref{FigCC}(b)-(c)].

\begin{figure}[t!]
\centering
\includegraphics[width=\columnwidth]{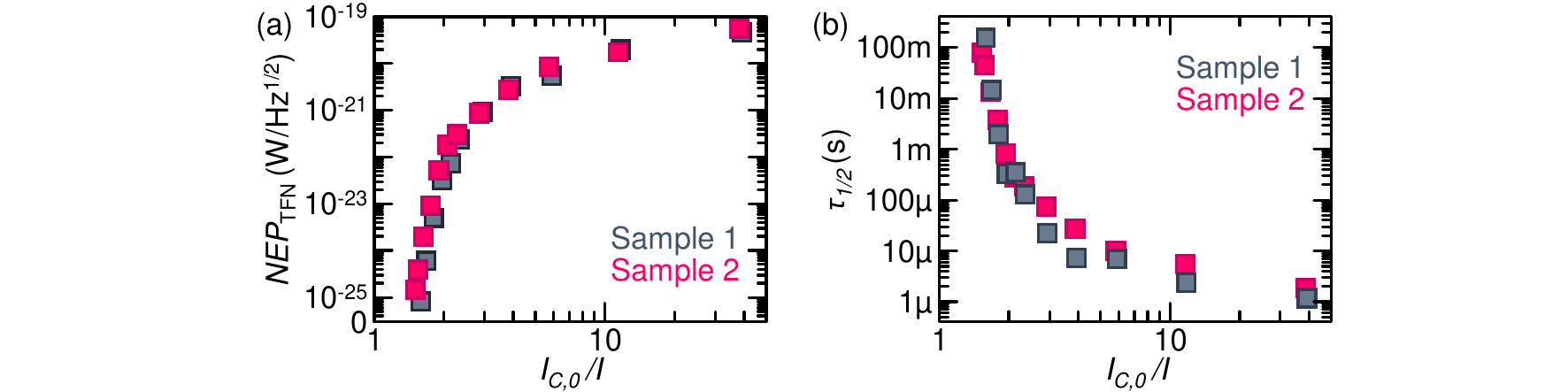}
\caption{Deduced properties of a JES. (\textbf{a}) Thermal fluctuations limited NEP as a function of the ratio $I_{C,0}/I$ for two different 1D-JJs. The NEP improves of about six orders of magnitude by increasing $I$. (\textbf{b}) JES bolometer time constant ($\tau_{1/2}$) as a function of the ratio $I_{C,0}/I$ for two different 1D-JJs. The high current operation decreases the speed of the bolometer.}
\label{FigJES}
\end{figure}

As expected, $NEP_{TFN,JES}$ can be in situ finely controlled by tuning $I$. In particular, it decreases monotonically by increasing the amplitude of bias current, as shown in Fig. \ref{FigJES}(a). Notably, the current bias modulates the device sensitivity over about 6 orders of magnitude.
Furthermore, the extracted data indicate that the JES \cite{Paolucci} reaches best sensitivities several orders of magnitude better than nano-TES bolometers \cite{Paolucci2} and all other sensors proposed and realized \cite{Irwin1995,Irwin2006,Virtanen,Kokkoniemi,Monfardini,Lee,Vischi,Brien,kuzmin2019,Rogalski,Ellrich,Wei,Khosropanah,Karasik,Lee1,Visser,Sun,Giazotto2008} so far. 
Specifically, the JES showed unprecedented values of $NEP_{TFN,JES}$ as low as $\sim 1\times 10^{-25}$ W/$\sqrt{\text{Hz}}$ for the highest bias current $I=370$ nA ($I_{C,0}/I\simeq1.5$) corresponding to a working temperature of about $18$ mK. This sensitivity arises from the extremely suppressed heat exchange of $A$ with all the thermal sinks (the phonons and the lateral electrodes) due to the combination of very low working temperature and operation deeply in the superconducting state. 

The JES time constant ($\tau_{1/2}$) can be calculated by the combination of Eqs. \ref{IntrinsicTime} and \ref{TauMezzi}. As for the TES, it is proportional to the ratio between the electron heat capacitance and the electron-phonon heat conductance in $A$ \cite{Giazotto2006}. Figure \ref{FigJES}(a) shows the dependence of the JES time constant deduced from the experimental data on the amplitude of $I$. As expected, $\tau_{1/2}$ is monotonically suppressed by rising $I$, since the thermalization of $A$ is exponentially suppressed by decreasing the working temperature. In particular, $\tau_{1/2}$ varies between $\sim 1\,\mu$s at $I_{C,0}/I\simeq42$ and $\sim 100\,$ms at $I_{C,0}/I\simeq1.5$.

The tunability of the JES properties allows to choose between moderate sensitivity/fast response and extreme sensitivity/slow response by simply varying the bias current within the same structure. As a consequence, the same bolometer could fulfill the requirements of different applications. 

\section{Conclusions}
\label{Concl}
This paper reviews a new class of superconducting bolometers 
owing the possibility of being finely in situ tuned by a bias current. This family of sensors includes the nanoscale transition edge sensor (nano-TES) \cite{Paolucci2} and the Josephson escape sensor (JES) \cite{Paolucci}, which take advantage of the strong resistance variation of a superconductor when transitioning to the normal-state. These bolometers employ a one dimensional fully superconducting Josephson junction (1D-JJ) as active region. On the one hand, this enables the modulation of the $R(T)$ characteristics of the device by varying $I$. On the other hand, the lateral superconducting electrodes ensure exponentially suppressed thermal losses, the so-called Andreev mirrors effect. 

The 1D-JJ is theoretically analyzed from an electronic and thermal transport point of view. In particular, the current dependence of the resistance versus temperature characteristics and of the electro-phonon thermalization are predicted. Furthermore, the equations of the main figures of merit of a bolometer, such as the noise equivalent power and response time are calculated both in the nano-TES and JES operations.

A complete series of electronic and thermal experiments allowed extracting all the parameters of the device active region, such as the current-dependent $R(T)$ characteristics, the density of states and the electron-phonon thermal relaxation. 
These data allow determining the performance of the sensing element the nano-TES and the JES while irradiated with continuous radiation, that is as a bolometer.
The nano-TES active region reaches a total noise equivalent power of about $5 \times 10^{-20}$ W$/\sqrt{\text{Hz}}$ limited by thermal fluctuations, thus suggesting the possibility to produce detectors with higher sensitivity than state-of-the-art conventional transition edge sensors \cite{Khosropanah}. In addition, the negative electrothermal feedback is expected to ensure a very fast sensing response $\tau_{eff}\simeq 10$ ns.
The JES showed the possibility of in situ tuning the sensitivity by changing the biasing current. The intrinsic thermal fluctuation limitation to NEP can be lowered down to the value of $\sim 1\times 10^{-25}$ W/$\sqrt{\text{Hz}}$. On the contrary, the sensing speed decreases significantly (up to about 100 ms), due to the poor thermalization of the active region when operated fully in the superconducting state. Thus, the JES might be the basis for unprecedented sensitive bolometers operating in the GHz range.

The nano-TES and the JES are expected to have a strong impact on radio astronomy \cite{Primack,Rowan2009}. In particular, studies of the CMB \cite{Seljak,Kamionkowski,Sironi}, CIB \cite{Cooray,Zemcov}, AGNs \cite{Fabian}, galaxies \cite{Villaescusa-Navarro,Armus}, comets \cite{Falchi1988}, gigahertz-peaked spectrum radio sources \cite{Odea1998}, supermassive black holes \cite{Issaoun2019} and radio  burst  sources \cite{Marcote2020} need new and hyper-sensitive radiation sensors operating in the gigahertz range. Furthermore, the nano-TES and the JES might be employed as well in imaging for medical purposes \cite{Sun}, quality controls in industrial plants \cite{Ellrich} and security scanners \cite{Rogalski}.


\vspace{6pt} 



\funding{This research was funded by the European Union's Horizon 2020 research and innovation programme under the grant No. 777222 ATTRACT (Project T-CONVERSE) and under FET-Open grant agreement No. 800923-SUPERTED, and by the CSN V of INFN under the technology innovation grant SIMP.}

\acknowledgments{We acknowledge P. Virtanen, P. Spagnolo, C. Gatti, R. Paoletti, F.S. Bergeret, G. De Simoni, E. Strambini, A. Tartari, G. Signorelli and G. Lamanna for fruitful discussions.}

\conflictsofinterest{The authors declare no conflict of interest.} 
\reftitle{References}



\begin{thebibliography}{999}
\bibitem{Primack}
Primack, J. R., Gilmore, R. C., Somerville, R. S., 
Diffuse Extragalactic Background Radiation, 
\emph{AIP Conference Proceedings} {\bf{2008}}, 1085, 71.

\bibitem{Seljak}
Seljak, U., and Zaldarriaga, M., 
Signature of Gravity Waves in the Polarization of the Microwave Background, 
\emph{Phys. Rev. Lett.} {\bf{1997}}, 78, 2054.

\bibitem{Sironi}
Sironi, G., The frequency spectrum of the Cosmic Microwave Background, 
\emph{New Astron. Rev.} {\bf{1999}}, 43, 243-249.

\bibitem{Kamionkowski}
Kamionkowski, M., and Kovetz, E. D., 
The Quest for B Modes from Inflationary Gravitational Waves, 
\emph{Annual Rev. Astron. and Astroph.} {\bf{2016}},  54, 227–269.

\bibitem{Tabatabaei2017}
Tabatabaei, F. S., et \textit{al.},
The Radio Spectral Energy Distribution and Star-formation Rate Calibration in Galaxies,
\emph{ApJ} {\bf{2017}}, 836, 185.

\bibitem{Fabian}
Fabian, A. C.,
Active galactic nuclei,
\emph{PNAS} {\bf{1999}}, 96, 4749-4751.

\bibitem{Cooray}
Cooray, A., et \textit{al.},
Near-infrared background anisotropies from diffuse intrahalo light of galaxies,
\emph{Nature} {\bf{2012}}, 490, 514–516.

\bibitem{Zemcov}
Zemcov, M., et \textit{al.},
On the origin of near-infrared extragalactic background light anisotropy,
\emph{Science} {\bf{2014}}, 346,  732-735.

\bibitem{Rowan2009}
Rowan-Robinson, M.,
Probing the Cold Universe,
\emph{Science} {\bf{2009}}, 325, 546-547.

\bibitem{Villaescusa-Navarro}
Villaescusa-Navarro, F., Planelles, S., Borgani, S., Viel, M., Rasia, E., Murante, G., Dolag, K., Steinborn, L. K., Biffi, V., Beck, A. M., and  Ragone-Figueroa, C.,
Neutral hydrogen in galaxy clusters: impact of AGN feedback and implications for intensity mapping, 
\emph{Monthly Notices of the Royal Astronomical Society} {\bf{2016}}, 456, 3553–3570.

\bibitem{Armus}
Armus, L., Charmandaris, V., and Soifer, B.T.,
Observations of luminous infrared galaxies with the Spitzer Space Telescope, 
\emph{Nat Astron} {\bf{2020}} 4, 467–477.

\bibitem{Marcote2020}
Marcote, B., et \textit{al.}
A repeating fast radio burst source localized to a nearby spiral galaxy,
\emph{Nature} {\bf{2020}}, 577, 190-194.

\bibitem{Falchi1988}
Falchi, A., Gagliardi, L., Palagi, F., Tofani, G., and Comoretto, G., 
\emph{10.7 GHz continuum observations of comet P/Halley} Springer: New York, USA, 1988.

\bibitem{Odea1998}
O'Dea, C. P.,
The Compact Steep-Spectrum and Gigahertz Peaked-Spectrum Radio Sources,
\emph{PASP} {\bf{1998}}, 110, 493-532.

\bibitem{Issaoun2019}
S. Issaoun, S., et \textit{al.},
The Size, Shape, and Scattering of Sagittarius A* at 86 GHz: First VLBI with ALMA,
\emph{ApJ} {\bf{2019}}, 871, 30.

\bibitem{Irwin2006}
Irwin, K. D., 
Seeing with Superconductors, 
\emph{Sci. Am.} {\bf{2006}}, 295, 86-94.

\bibitem{Irwin1995}
Irwin, K. D.,
An application of electrothermal feedback for high resolution cryogenic particle detection,
\emph{Appl. Phys. Lett.} {\bf{1995}}, 66, 1998-2000.

\bibitem{Karasik}
Karasik,B. S., and Cantor, R., 
Demonstration of high optical sensitivity in far-infrared hot-electron bolometer, \emph{Appl. Phys. Lett.} {\bf{2011}}, 98, 193503.

\bibitem{Monfardini} 
Monfardini, A., et \textit{al.},
Lumped element kinetic inductance detectors for space applications, 
\emph{Proc. SPIE 9914, Millimeter, Submillimeter, and Far-Infrared Detectors and Instrumentation for Astronomy VIII} {\bf{2016}}, 99140N.

\bibitem{Khosropanah}
Khosropanah, P., et \textit{al.},
Low noise transition edge sensor (TES) for the SAFARI Instrument on SPICA. 
\emph{Proc. SPIE 7741, Millimeter, Submillimeter, and Far-Infrared
Detectors and Instrumentation for Astronomy V} {\bf{2010}}, 77410L.

\bibitem{Visser}
de Visser, P., et \textit{al.},
Fluctuations in the electron system of a superconductor exposed to a photon flux, \emph{Nat Commun.} {\bf{2014}}, 5, 3130.

\bibitem{Wei}
Wei, J., Olaya, D., Karasik, B. S., Pereverzev, S. V., Sergeev,  A. V. and Gershenson, M. E.,
Ultrasensitive hot-electron nanobolometers for terahertz astrophysics, 
\emph{Nat Nanotech.} {\bf{2008}}, 3, 496–500.

\bibitem{Josephson}
Josephson, B. D.,
Possible new effects in superconductive tunnelling,
\emph{Phys. Lett.} {\bf{1962}}, 1, 251-253.

\bibitem{Solinas}
Solinas, P., Giazotto, F., and Pepe, G. P.,
Proximity SQUID Single-Photon Detector via Temperature-to-Voltage Conversion, 
\emph{Phys. Rev. Applied} {\bf{2018}}, 10, 024015.

\bibitem{Guarcello}
Guarcello, C., Braggio, A., Solinas, P., Pepe, G. P., and Giazotto, F., 
Josephson-Threshold Calorimeter, 
\emph{Phys. Rev. Applied} {\bf{2019}}, 11, 054074.

\bibitem{kuzmin2019}
Kuzmin, L. S., et al., 
Photon-noise-limited cold-electron bolometer based on strong electron self-cooling for high-performance cosmology missions, 
\emph{Commun Phys} {\bf{2019}}, 2, 104.

\bibitem{Brien}
Brien, T. L. R., et \textit{al.},
A strained silicon cold electron bolometer using Schottky contacts, 
\emph{Appl. Phys. Lett.} {\bf{2014}}, 105, 043509.

\bibitem{Vischi}
Vischi, F., Carrega, M., Braggio, A., Paolucci, F., Bianco, F., Roddaro, S., and Giazotto, F., 
Electron Cooling with Graphene-Insulator-Superconductor Tunnel Junctions for Applications in Fast Bolometry, 
\emph{Phys. Rev. Applied} {\bf{2020}}, 13, 054006.

\bibitem{Giazotto2008}
Giazotto, F., Heikkil\"a, Pepe, G. P., Helist\"o, P., T. T., Luukanen, and Pekola, J. P.,
Ultrasensitive proximity Josephson sensor with kinetic inductance readout, 
\emph{Appl. Phys. Lett.} {\bf{2008}}, 92, 162507.

\bibitem{Kokkoniemi}
Kokkoniemi, R., et \textit{al.},
Nanobolometer with ultralow noise equivalent power,
\emph{Commun Phys} {\bf{2019}}, 2, 124.

\bibitem{Lee1}
Lee, G.-H., et \textit{al.},
Graphene-based Josephson junction microwave bolometer,
\emph{Nature} {\bf{2020}}, 586, 42–46.

\bibitem{Virtanen}
Virtanen, P., Ronzani, A., and Giazotto, F.,
Josephson Photodetectors via Temperature-to-Phase Conversion, 
\emph{Phys. Rev. Applied} {\bf{2018}}, 9, 054027.

\bibitem{Sun}
Sun, Q., He, Y., Liu, K., Fan, S., Parrott, E. P. J., and Pickwell-MacPherson,  E.,
Recent Advances in Terahertz Technology for Biomedical Applications, 
\emph{Quant. Imaging Med. Surg.} {\bf{2017}}, 7, 345-355.

\bibitem{Ellrich}
Ellrich, F., et \textit{al.},
Terahertz Quality Inspection for Automotive and Aviation Industries,  
\emph{J. Infrared Milli. Terahz. Waves} {\bf{2020}}, 41, 470–489.

\bibitem{Rogalski}
Rogalski, A., and Sizov, F.,
Terahertz detectors and focal plane arrays, 
\emph{Opto-Electronics Review}, {\bf{2011}}, 19, 346–404.

\bibitem{Paolucci2}
Paolucci, F., Buccheri, V., Germanese, G., Ligato, N., Paoletti, R., Signorelli, G., Bitossi., M., Spagnolo, P., Falferi, P., Rajteri, M., Gatti, C., and Giazotto, F.,
Highly sensitive nano-TESs for gigahertz astronomy and dark matter search
\emph{J. Appl. Phys.}, {\bf{2020}}, in press.

\bibitem{Paolucci}
Paolucci, F., Ligato, N., Buccheri, V., Germanese, G., Virtanen, P., and Giazotto, F.,
Hypersensitive tunable Josephson escape sensor for gigahertz astronomy, 
\emph{Phys. Rev. Applied}, {\bf{2020}}, 14, 034055.

\bibitem{tinkham}
Tinkham, M., \emph{Introduction to Superconductivity}, McGraw-Hill: New York, USA, 1996.

\bibitem{Barone1982}
Barone, A., and Patern\`o, G.,
\emph{Physics and Applications of the Josephson Effect}, Wiley-VCH: New York, USA, 1982.

\bibitem{Bezryadin2012}
Bezryadin, A.,
\textit{Superconductivity in Nanowires: Fabrication and Quantum Transport} Wiley-VCH: New York, USA, 2012.  

\bibitem{Zant1992}
van der Zant, H. J. S.,  Fritschy, F. C., Elion, W. J., Geerlings, L. J., and Mooij, J. E.,
Field-Induced Superconductor-to-Insulator Transitions in Josephson-Junction Arrays,
\emph{Phys. Rev. Lett.} {\bf{1992}}, 69, 2971-2974.

\bibitem{Ivanchenko1969}
Ivanchenko, Yu. M., and Zil'berman, L. A.,
The Josephson effect in small tunnel contacts,
\emph{Sov. Phys. JETP} {\bf{1969}}, 28, 1272.

\bibitem{Andreev1964}
Andreev, A. F.,
The Thermal Conductivity of the Intermediate State in Superconductors,
\emph{JETP} {\bf{1964}}, 66, 1228-1231.

\bibitem{Giazotto2006}
Giazotto, F., Heikkil\"a, T. T., Luukanen, A., Savin, A. M., and Pekola, J. P.,
Opportunities for mesoscopics in thermometry and refrigeration: Physics and applications,
\emph{Rev. Mod. Phys.} {\bf{2006}}, 78, 217-274.

\bibitem{Timofeev2009}
Timofeev, A. V., Pascual Giarc\'ia, C., Kopnin, N. B., Savin, A. M., Meschke, M., Giazotto, F., and Pekola, J. P.,
Recombination-Limited Energy Relaxation in a Bardeen-Cooper-Schrieffer Superconductor,
\emph{Phys. Rev. Lett.} {\bf{2009}}, 102, 017003.

\bibitem{Courtois}
Courtois, H., Meschke, M., Peltonen, J. T., and Pekola, J. P.,
Hysteresis in a Proximity Josephson Junction,
\emph{Phys. Rev. Lett.} {\bf{2008}}, 101, 067002.

\bibitem{Mather}
Mather, J. C., 
Bolometer noise: non equilibrium theory, 
\emph{Appl. Opt.} {\bf{1982}}, 21, 1125-1129.

\bibitem{Lee}
Lee, S., Gildemeister, J. M., Holmes,  W., Lee,  A. T., and Richards, P. L., Voltage-biased superconducting transition-edge bolometer with strong electrothermal feedback operated at 370 mK, 
\emph{Appl. Opt.} {\bf{1998}}, 37, 3391-3397.

\bibitem{Bergmann}
Bergmann, T.,
\emph{Energy resolving power of transition edge x-ray microcalorimeters}. 
Phd Dissertation University of Utrecht, Utrecht, Netherlands, 2004.

\bibitem{Heikkila}
Heikkil$\ddot{a}$, T. T., Silaev, M., Virtanen, P., and Bergeret, F. S.,
Thermal, electric and spin transport in superconductor/ferromagnetic-insulator structures. 
\emph{Progress in Surface Science} {\bf{2019}}, 94, 100540.,

\bibitem{Rabani}
Rabani, H., Taddei, F., Bourgeois, O., Fazio, R., and Giazotto, F.,
Phase-dependent electronic specific heat in mesoscopic josephson junction. \emph{Phys. Rev. B} {\bf{2008}}, 78. 012503. 

\bibitem{cochran}
Cochran, J. F.,  and Mapother, D. E.,
Superconducting Transition in Aluminum, 
\emph{Phys Rev.} {\bf{1958}}, 111, 132.

\bibitem{DeGennes1964}
De Gennes, P. G.,
Boundary Effects in Superconductors,
\emph{Rev. Mod. Phys.} {\bf{1964}}, 36, 225.

\bibitem{Kogan1982}
Kogan, V. G.,
Coherence length of a normal metal in a proximity system,
\emph{Phys. Rev. B} {\bf{1982}}, 26, 88.

\bibitem{Golubov}
Golubov, A. A., Kupriyanov, M. Yu., and Il’ichev, E.,
The current-phase relation in Josephson junctions,
\emph{Rev. Mod. Phys.} {\bf{2004}}, 76, 411-479.

\end{thebibliography}




\end{document}